\documentstyle[spie]{article} 
\input{psfig}   

\title{Probing black hole X-ray binaries with the Keck telescopes}

\author{Emilios T. Harlaftis\supit{a} and Alexei V. Filippenko\supit{b} 
\skiplinehalf 
\supit{a}Institute of Astronomy and Astrophysics, National Observatory \\
of Athens, P. O. Box 20048, Athens - 118 20, Greece\\
\supit{b}Department of Astronomy, University of California,
                Berkeley, CA 94720-3411, USA\\
}

\authorinfo{Further author information: (Send correspondence to 
E. T. Harlaftis)\\E. T. Harlaftis: E-mail: ehh@astro.noa.gr\\ 
A. V. Filippenko: E-mail: alex@astro.berkeley.edu}

\begin{document} 
  \maketitle 

\begin{abstract}
The advent of the large effective apertures of the Keck telescopes has
resulted in the determination with  unprecedented accuracy of the mass
functions  and  mass ratios of    faint ($R  \approx 21$  mag)   X-ray
transients (GS~2000+25,  GRO~J0422+32, Nova Oph 1977, Nova Vel 1993),
as well as constraining the main-sequence companion star parameters and 
producing images of the accretion disks around the black holes. 
\end{abstract}



\section{INTRODUCTION}
\label{sect:intro}  

Zel'dovich and Novikov (1966) were the first  to propose the technique
which is  still in use  for  ``weighing" black holes.  They  suggested
that  black holes  could  be detected   indirectly  from light emitted
through the interaction with  a donor star  in an X-ray binary system.
The motion of  the donor star  around the black  hole  would produce a
radial velocity  sinusoidal curve  which  could be detected   from the
Doppler shifts of the photospheric absorption lines of the donor star.
The semi-amplitude ($K$) of the curve  together with the binary period
($P$) determine the mass function of the black  hole (a lower limit to
its mass), using Kepler's third law: $f_{x} = PK^3/(2\pi G)$.  Indeed,
X-ray binaries were  found in the late 1960s  and the first black-hole
candidate, Cyg X-1, in 1971 (Oda et  al.  1971).  Efforts in measuring
the mass of Cyg X-1 were affected by uncertainties in
the evolution of the massive donor star, and with  a low mass function
($f_{x}=0.22\pm0.01~M_{\odot}$; Bolton  1975) this was not regarded as
unequivocal evidence for  a black hole (see Herrero  et al.  1995  for
the most recent work).

\section{HUNTING FOR BLACK HOLES IN X-RAY NOVAE}
 
In the 1980s the observational effort was turned to X-ray novae (XRNs,
a sub-group of low-mass X-ray binaries).  Unlike classical novae, XRNs
are  accretion-driven events that show  disk  outbursts with a typical
rise of 8--10 mag in a few days and  a subsequent decline over several
months.  After  the  XRN has  subsided into quiescence,  the accretion
disk does not dominate the observed flux, rendering the companion star
visible.  The low-mass companion star allows the  mass function of the
black hole, a  good approximation of  the  mass in  a high-inclination
system, to be determined.  In the 1990s, X-ray satellites found 6 XRNs
with identified companion  stars  in the  optical (Nova Muscae   1991,
Cheng  et al.   1992; Nova  Persei 1992,   Casares  et al.   1995a and
references therein; Nova Sco 1994, Bailyn et al.  1995; Nova Vel 1993,
Filippenko  et  al.  1999;  GRO J1719-24,  e.g., Ballet et  al.  1993;
XTE~J1550-564, e.g., Smith et al. 1998).

The prototype target in the 1980s was A0620--00, but unfortunately its
mass function   was close  to  the  maximum  mass of   a neutron  star
($f_{x}=3.2\pm0.2~M_{\odot}$; McClintock and  Remillard 1986).  It was
not until 1992 that a mass function of a candidate black hole in the XRN
1989, GS~2023+338, was found to be much heavier  than the maximum mass
of  a    neutron  star ($f_{x}=6.08\pm0.06~M_{\odot}$;    Casares   et
al.  1992).  Since    then, efforts  have    been  directed toward
measuring actual  masses,   thus  producing the  first  observed  mass
distribution of black holes (Bailyn et al.  1998; Miller et al.  1998;
for  the theoretical distribution see  Fryer 1999).  The determination
of   the masses  of  stellar remnants  after  supernova explosions  is
essential for  an understanding of   the late  stages of evolution  of
massive stars.  Very recently,   the first observational  evidence for
the progenitor, a supernova or hypernova with a mass $> 30 M_{\odot}$,
that produced the black hole of $7.0\pm0.2 ~M_{\odot}$ in GRO~J1655-40
was found with  the Keck-I telescope  (from high metal abundances that
were presumably deposited onto the surface  of the companion F5IV-star
by the supernova explosion; Israelian et al.  1999).

   \begin{figure*}
   \begin{center}
   \begin{tabular}{c}
   \psfig{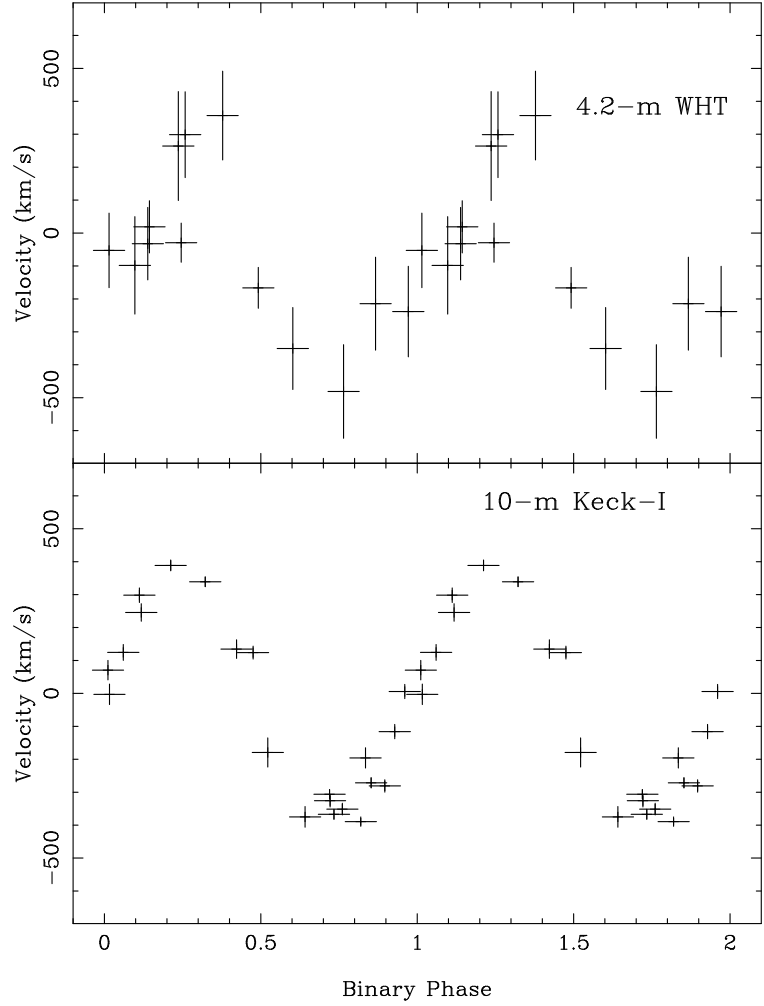} 
   \end{tabular}
   \end{center}
   \caption[example] 
   { \label{fig:example}	  
{\it Bottom:} the radial velocity curve of the companion   
star to  the black hole   GRO~J0422+32 as extracted from spectra  near
H$\alpha$ with Keck-I/LRIS (Harlaftis et al. 1999).  {\it
Top:} the radial  velocity curve of the  companion  star to the  black
hole  GRO~J0422+32  as  extracted from  4.2-m  WHT/ISIS  near-infrared
spectra (8450-8750~\AA)    (Casares et al.  1995a).  The
reduction in  the individual measurement  uncertainties is a factor of
four using Keck-I.  The sinusoidal fit  to the radial velocities gives
$K  = 338\pm39$ km s$^{-1}$  with the WHT  data  and $K = 372\pm10$ km
s$^{-1}$ with the Keck data for the  radial velocity semi-amplitude of
the companion star.   This yields better  accuracy in  the estimate of
the lower limit of the  black hole's mass, from $P K^3 /(2 \pi G) =
0.85\pm0.30~M_{\odot}$      to   $1.13\pm0.09~M_{\odot}$      for   the
low-inclination system GRO~J0422+32. WHT data courtesy of Jorge Casares.} 
   \end{figure*}

\section{RADIAL VELOCITY CURVES}

Utilizing the  Doppler effect produced   by the shifting  photospheric
lines due to the orbital motion of the companion star around the black
hole, but now with the 10-m Keck-I  and Keck-II telescopes, Filippenko
and his collaborators have produced the four
most accurate mass functions ($f_{x}=5.0\pm0.1
~M_{\odot}$,      $1.2     \pm0.1~M_{\odot}$,   $4.7\pm0.2~M_{\odot}$,
$3.2\pm0.1~M_{\odot}$, respectively for GS~2000+25, GRO~J0422+32, Nova
Oph 1977, Nova Vel 1993; Filippenko et al. 1995a, 1995b, 1997, 1999).  
Figures 1 and  2 show the great improvement
that the large aperture of Keck offers in comparison to 4-m-class
telescopes in extracting radial  velocity curves of  the motion of the
donor star around  the  black hole by  cross-correlating main-sequence
template spectra with the observed spectra.

\section{THE MAIN SEQUENCE COMPANION STAR}

The  line broadening  function affecting  the  absorption lines of the
object spectra consists of the convolution of the instrumental profile
(full width at half-maximum = 108 km s$^{-1}$) with the companion star's 
rotational broadening profile (of width $\upsilon \sin i$), 
with further smearing  due  to changes in the  orbital
velocity of the companion star during  a given exposure.  The exposure
time  for  each object  spectrum  ($T_{\rm  exp}~\approx$ 25--40  min)
resulted in orbital smearing of the lines up to $2\pi K_{\rm c} T_{\rm
exp}/P$, which can range up to 242  km  s$^{-1}$; hence,   the template
spectra were subsequently smeared by  the amount corresponding to  the
orbital motion through convolution  with a rectangular profile and the
resulting  template spectrum was  further broadened from  2  to 150 km
s$^{-1}$  by convolution with the  Gray (1976) rotational profile.  We
scaled the blurred template spectrum by a factor $0  < f < 1$ to match
the  absorption-line strengths    in the  Doppler-corrected    average
spectrum.  Finally, the simulated template spectrum (i.e., smeared and
broadened) was subtracted from  the Doppler-corrected average spectrum
of Nova Oph 1977  and  $\chi^{2}$ was computed from a smoothed version
of the  residual spectrum.  The minimum  $\chi^{2}$  gives the optimal
$\upsilon \sin  i$, $f$, and spectral  type of the  companion star (for
more details see Harlaftis et al.  1996, 1997, 1999).

   \begin{figure*}
   \begin{center}
   \begin{tabular}{c}
   \psfig{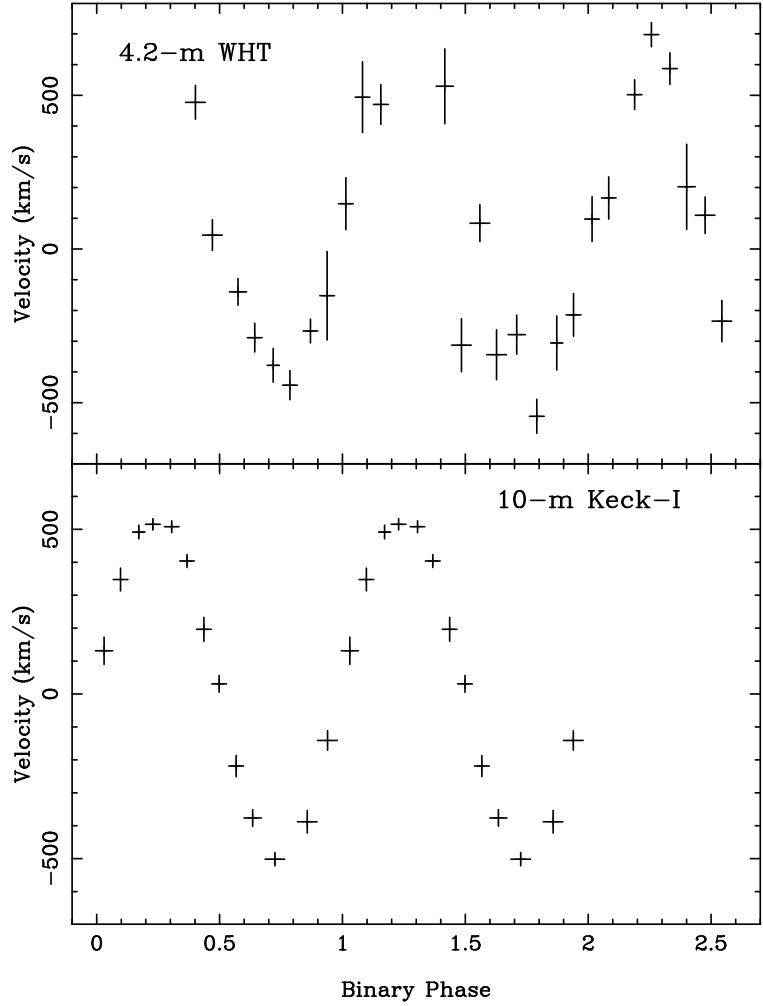} 
   \end{tabular}
   \end{center}
   \caption[example] 
   {  \label{fig:example} {\it Bottom:} the   radial velocity curve of
   the companion star  to the black  hole GS~2000+25 as extracted from
   spectra near H$\alpha$ with  Keck-I/LRIS in just 1 night (Harlaftis
   et  al. 1996).     {\it Top:}  the radial   velocity  curve of  the
   companion star  to the  black hole  GRO~J0422+32  as extracted from
   4.2-m WHT/ISIS in 3 nights (Casares  et al.  1995b). The sinusoidal
   fit to the radial velocities gives $K =  520\pm16$ km s$^{-1}$ with
   the WHT data and  $K = 520\pm5$ km s$^{-1}$  with the Keck data for
   the  radial velocity  semi-amplitude  of the  companion star.  This
   yields better accuracy  in the estimate of the  lower limit  of the
   black      hole's      mass,    from     $PK^{3}/(2\pi G)     =
   5.02\pm0.46~M_{\odot}$   to    $5.01\pm0.15~M_{\odot}$   for     the
   high-inclination system   GS~2000+25. WHT  data courtesy  of  Jorge
   Casares.}  \end{figure*}

Figure   3 summarizes  the procedure  we    follow to deconvolve   the
main-sequence spectrum from the  target spectrum.  The spectrum of a
M2~V template (BD~$+44^{\circ}2051$) is shown at the bottom, binned to
124 km s$^{-1}$  pixels (=4  pixels  and similar to the   instrumental
resolution).  This template was then treated so that its line profiles
simulate those  of  the GRO~J0422+32  spectra. The smearing  in radial
velocity  due to the orbital  line broadening while exposing are applied to
individual copies of   the M2 template,  and  these were  subsequently
averaged using  weights  identical  to   those corresponding to    the
GRO~J0422+32   spectra.  Next,    a   rotational  broadening   profile
corresponding  to $\upsilon \sin i =  50$ km s$^{-1}$ was applied; the
result is the   second spectrum  from the  bottom  in  Figure 3.   The
spectrum above  is the Doppler-corrected average of the GRO~J0422+32
data in the rest  frame of the  M2~V template.  Finally, the  residual
spectrum is  shown at  the top  after 0.61  times  the simulated  M2~V
template ($f = 0.61 \pm 0.04$ for M2; Table 4) was subtracted from the
Doppler-shifted average spectrum.  The M-star absorption lines and TiO
bands are evident  in   the Doppler-corrected  average, and   they are
almost  completely removed by   subtraction  of the  template spectrum
(e.g., the  Na~I~D line).  Emission  from He{\small~I}   $\lambda$5876
becomes  prominent after  subtraction of  the  M2~V template  and weak
emission  from He{\small~I} $\lambda$6678  is  also present. Note that
there is no evidence for Li{\small~I}  $\lambda$6708 absorption, to an
equivalent width upper  limit  of 0.13~\AA\  ($1\sigma$)  relative  
to the  original continuum, except in GS~2000+25 (see Mart\'\i n
et al.  1994 for lithium in X-ray binaries).

   \begin{figure*}
   \begin{center}
   \begin{tabular}{c}
   \psfig{figure=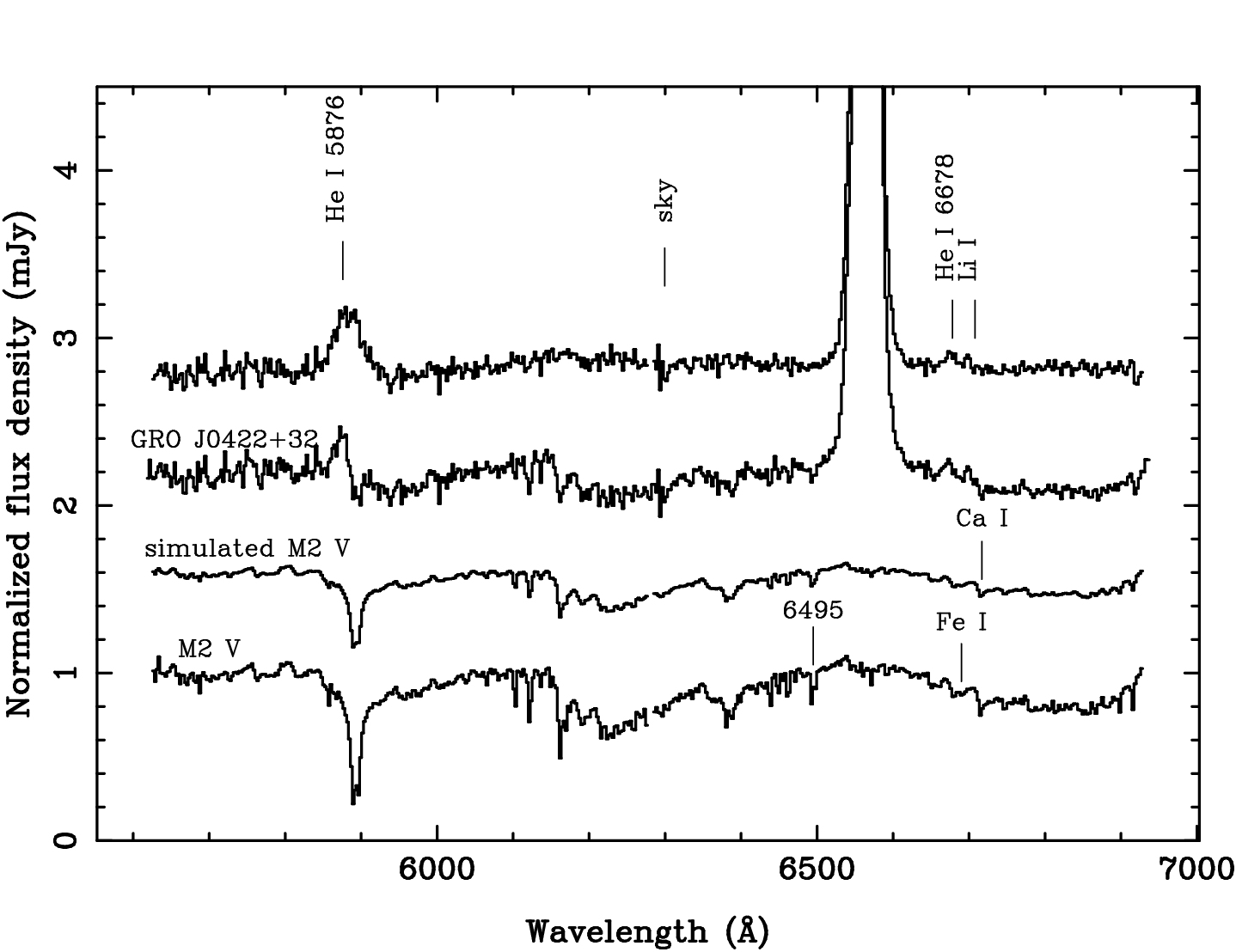,height=12cm,angle=-90} 
   \end{tabular}
   \end{center}
   \caption[example] 
   { \label{fig:example}	  

Results of  the technique followed to  extract  $\upsilon \sin i$, $f$,
and the spectral type of the companion star.  From bottom to top: the M2~V
template  BD +$44^{\circ}2051$,   the M2~V template  convolved  with a
complex profile to simulate effects of orbital smearing and rotational
broadening  ($\upsilon \sin i =  50$ km s$^{-1}$), the Doppler-shifted
average  spectrum  of GRO~J0422+32,   and  the  residual spectrum   of
GRO~J0422+32  after  subtraction of  the M2~V template times $f  =
0.61$. The spectra are binned to 124 km  s$^{-1}$ pixels. An offset of
0.6  mJy was   added to  each   successive spectrum for clarity.   The
residual  spectrum   is dominated by the   disk  spectrum (e.g., broad
H$\alpha$ and He{\small~I} lines in emission). Several other lines are
also  marked,  such as   the  characteristic Fe{\small~I}+Ca{\small~I}
blend at   6495~\AA\  in G--M stars,   as well  as   the Ca{\small~I}
6717~\AA\ and  Fe{\small~I} lines surrounding the  absent Li{\small~I}
line at 6707.8~\AA.  }
\end{figure*}

\section{THE MASS RATIO OF THE BLACK HOLE BINARIES}

Determination  of  the mass   ratio  (from  the  rotational
broadening of the  photospheric lines in the  companion  star) and the
inclination   (inferred  from   the ellipsoidal    modulations  of the
companion star), when  combined  with  the mass function,   can  fully
describe the  system's   parameters  and  the  masses of  the   binary
components. The mass ratio $q  = M_{2}/M_{1}$  is found by measuring
the  rotational broadening of the  absorption  lines of the companion,
$\upsilon \sin i$, through the relation

\[ \frac{\upsilon \sin i}{K_{c}} = 
0.46 \left[ (1+q)^{2} ~q \right] ^{1/3}, \]

\noindent
which is valid since the binary period  is so short that the companion
star is tidally  locked to the  black hole. We determined mass ratios 
for the first time for binaries as faint as 21  mag using the 
$\chi^{2}$ optimization   technique described in the
previous  section to  extract  the   rotational broadening of   the
absorption lines of  the donor star (Harlaftis et al. 1996, 1997, 1999).
The complete results of 
the analysis of the Keck data are given in Table 1.

\begin{table} [h]
\caption{Keck-deduced parameters}
\renewcommand{\arraystretch}{1.4}
\setlength\tabcolsep{5pt}
\begin{center}
\begin{tabular}{|l|l|l|l|l|}
\hline\noalign{\smallskip}
&Oph 1977 & GRO~J0422+32 & GS~2000+25   & Vel 1993\\
\hline
\rule[-1ex]{0pt}{3.5ex}  $K_{c}$ (km s$^{-1}$)&441$\pm$6 &372$\pm$10     &520$\pm$5      &475$\pm$6\\
\hline
\rule[-1ex]{0pt}{3.5ex}  $f_{x}$ &4.65$\pm$0.21  &1.13$\pm$0.09  &5.01$\pm$0.15  &3.17$\pm$0.12\\
\hline
\rule[-1ex]{0pt}{3.5ex}  Spectral type& K5V$\pm$2&M2V$^{+2}_{-1}$        &K5V$^{+1}_{-2}$&K8V$\pm$2\\
\hline
\rule[-1ex]{0pt}{3.5ex}  $f$ \%  & 30$\pm$3      & 61$\pm$4              & 94$\pm$5      &\\
 \hline
\rule[-1ex]{0pt}{3.5ex} 
$\upsilon \sin$~i (km s$^{-1}$)& $50^{+17}_{-23}$&90$^{+22}_{-27}$&86$\pm$8&\\
\hline
\rule[-1ex]{0pt}{3.5ex}  $q$&0.014$^{+0.019}_{-0.012}$ & 0.116$^{+0.079}_{-0.071}$ & 0.042$\pm$0.012 &\\
\hline
\end{tabular}
\end{center}
\end{table}

\section{THE ACCRETION DISK}

The  accretion disk in its  quiescent state has mainly been undetected
so far by X-ray satellites but can be studied in  the optical.  
Double-peaked  
Balmer profiles are  observed  with ``S''-wave components either
from the    companion   star (Nova Oph 1977)      or the bright   spot
(GS~2000+25)  and an H$\alpha$ emissivity law is observed, 
similar to that seen in 
dwarf novae.  An   imaging technique, Doppler  tomography,  shows the
accretion disks in  GS~2000+25, Nova Oph  1977 and  GRO~J0422+32 to be
present (Fig. 4).   
Further,    mass transfer from    the donor  star continues
vigorously to the outer disk as  evidenced by the ``bright spot,'' the
impact of the gas  stream onto the  outer accretion disk in GS~2000+25
(Fig. 4; Harlaftis et al.   1996).  

   \begin{figure*}
   \begin{center}
   \begin{tabular}{c}
   \psfig{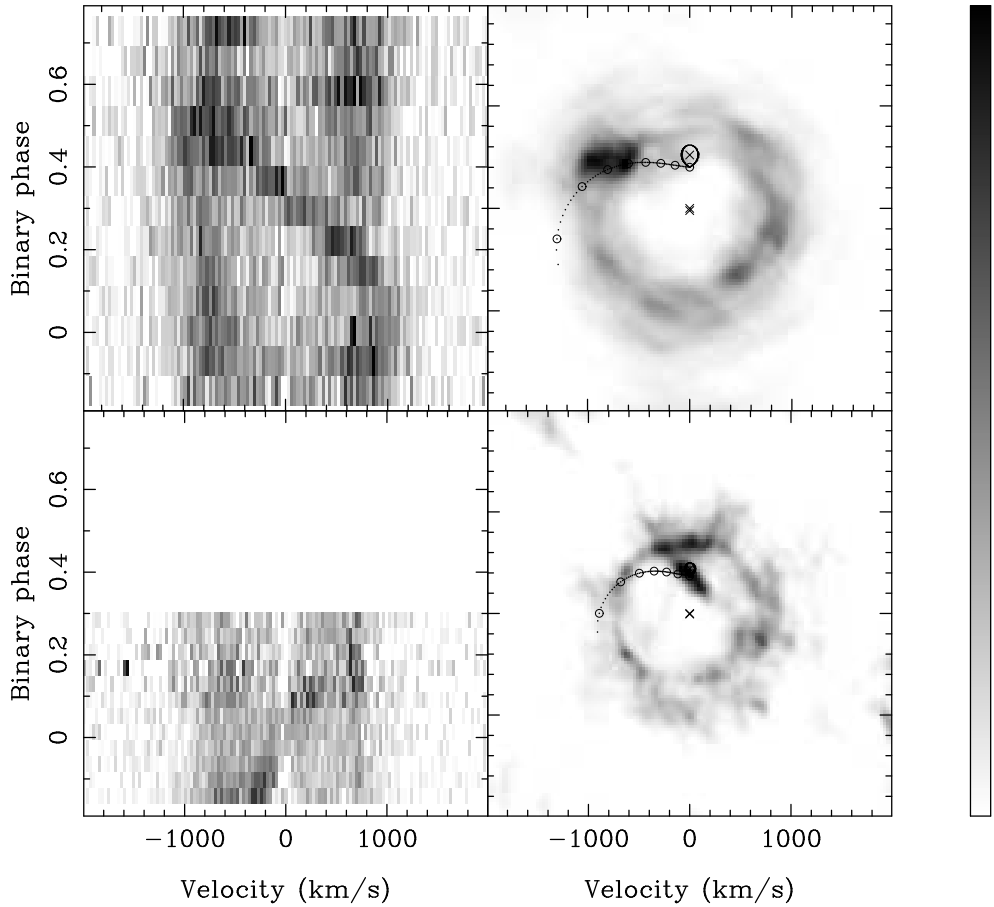} 
   \end{tabular}
   \end{center}
   \caption[example] 
   { \label{fig:example} The H$\alpha$  Doppler image  ({\it top-right
   panel}) of the accretion disk surrounding the black hole GS~2000+25
   ({\it bottom-right panel} for Nova Oph 1977), as reconstructed from
   13   Keck-I/LRIS spectra which   are also  presented ({\it top-left
   panel}; {\it  bottom-left  panel} for the 12   spectra of  Nova Oph
   1977).  By projecting   the image in   a  particular direction, one
   obtains  the  H$\alpha$   emission-line profile  as a   function of
   velocity;  for  example, projecting  toward the  top results in the
   profile  at orbital phase 0.0,  which has  a blueshifted peak. The
   path in velocity  coordinates of gas  streaming  from the dwarf  K5
   secondary star is illustrated.  The  GS~2000+25 Doppler map shows a
   bright spot,  at the    upper  left quadrant, which  results    from
   collision of the   gas stream with the  accretion  disk around  the
   black hole. The  Nova  Oph 1977 map  also  shows a trace of an ``S''-wave
   component  which, however, is  not resolved with clarity. The image
   was reconstructed by applying Doppler tomography, a maximum entropy
   technique, to the phase-resolved spectra, as described by Harlaftis
   et al. (1996, 1997, 1999). } \end{figure*}




\begin{thebibliography}{99}   

\bibitem{}
Bailyn, C., et al. (1995) Nature, 378, 157

\bibitem{}
Bailyn, C. D., Jain, R. K., Coppi, P., Orosz, J. A. (1998) ApJ, 499, 367

\bibitem{}
Ballet, J., et al. (1993) IAUC No. 5874

\bibitem{}
Bolton, C. T. (1975) ApJ, 200, 269

\bibitem{}
Casares, J., et al. (1992) Nature, 355, 614

\bibitem{}
Casares, J., et al. (1995a) MNRAS, 276, 35 

\bibitem{}
Casares, J., et al. (1995b) MNRAS, 277, L45

\bibitem{}
Cheng, F. H., et al. (1992) 397, 664


\bibitem{}
Filippenko, A. V., Matheson, T., Barth, A. J. (1995a) ApJ, 455, L139


\bibitem{}
Filippenko, A. V., Matheson, T., Ho, L. C. (1995b) ApJ, 455, 614

\bibitem{}
Filippenko, A. V., et al. (1997) PASP, 109, 461

\bibitem{}
Filippenko, A. V., et al. (1999) PASP, 111, 969

\bibitem{}
Fryer, C. L. (1999) ApJ, 522, 413

\bibitem{}
Gray, D. F. (1976) The Observations and Analysis of Stellar Photospheres (New
   York: Wiley-Interscience), p. 373

\bibitem{}
Harlaftis, E. T., Collier, S. J., Horne, K., Filippenko, A.  V. (1999) A\&A, 341, 491

\bibitem{}
Harlaftis, E. T., Horne, K., Filippenko, A. V. (1996) PASP, 108, 762

\bibitem{}
Harlaftis, E. T., Steeghs, D., Horne, K., Filippenko, A. V. (1997) AJ, 114, 1170

\bibitem{}
Herrero, A., et al. (1995), A\&A, 297, 556

\bibitem{}
Israelian, G., et al. (1999) Nature, 401, 142

\bibitem{}
McClintock, J. E., Remillard, R. A. (1986) ApJ, 308, 110 

\bibitem{}
Mart\'\i n, E., et al. (1994) ApJ,  435, 791

\bibitem{}
Miller, J. C., Shahbaz, T., Nolan, L. A. (1998) MNRAS, 294, L25

\bibitem{}
Oda, M., et al. (1971) ApJ, 166, L10

\bibitem{}
Smith, D. A., Marshall, F. E., Smith, E. A. (1998) IAUC No. 7008

\bibitem{}
Zel'dovich, Ya. B., Novikov, I. D. (1966) Sov. Physics -- Uspekhi, 8, 522
\end{thebibliography}
  \end{document}